\documentclass[sigconf,authorversion]{acmart}

\usepackage{siunitx}
\usepackage{colortbl}
\usepackage{listings}
\usepackage{framed}
\usepackage{enumitem}
\usepackage{subcaption}

\usepackage{xstring}

\newcommand{\pval}[1]{
  {\scriptsize(\textit{p}
    \IfBeginWith{#1}
    {<}{}{=}
    #1)
  }
}

\lstdefinestyle{monocode}{
  basicstyle=\ttfamily\small,
  columns=fullflexible,
  keepspaces=true,
  showstringspaces=false,
  breaklines=true,
  tabsize=4,
  language=Python,
  keywordstyle={}
}

\begin{document}
\title[Computer Science Achievement and Writing Skills Predict Vibe Coding Proficiency]{Computer Science Achievement and Writing Skills \\ Predict Vibe Coding Proficiency}

\begin{abstract}
Many software development platforms now support LLM-driven programming, or ``vibe coding'', a technique that allows one to specify programs in natural language and iterate from observed behavior, all without directly editing source code. While its adoption is accelerating, little is known about which skills best predict success in this workflow. We report a preregistered cross-sectional study with tertiary-level students (N = 100) who completed measures of computer-science achievement, domain-general cognitive skills, written-communication proficiency, and a vibe-coding assessment. Tasks were curated via an eight-expert consensus process and executed in a purpose-built, vibe-coding environment that mirrors commercial tools while enabling controlled evaluation. We find that both writing skill and CS achievement are significant predictors of vibe-coding performance, and that CS achievement remains a significant predictor after controlling for domain-general cognitive skills. The results may inform tool and curriculum design, including when to emphasize prompt-writing versus CS fundamentals to support future software creators.
\end{abstract}



\begin{teaserfigure}
 \includegraphics[width=\textwidth]{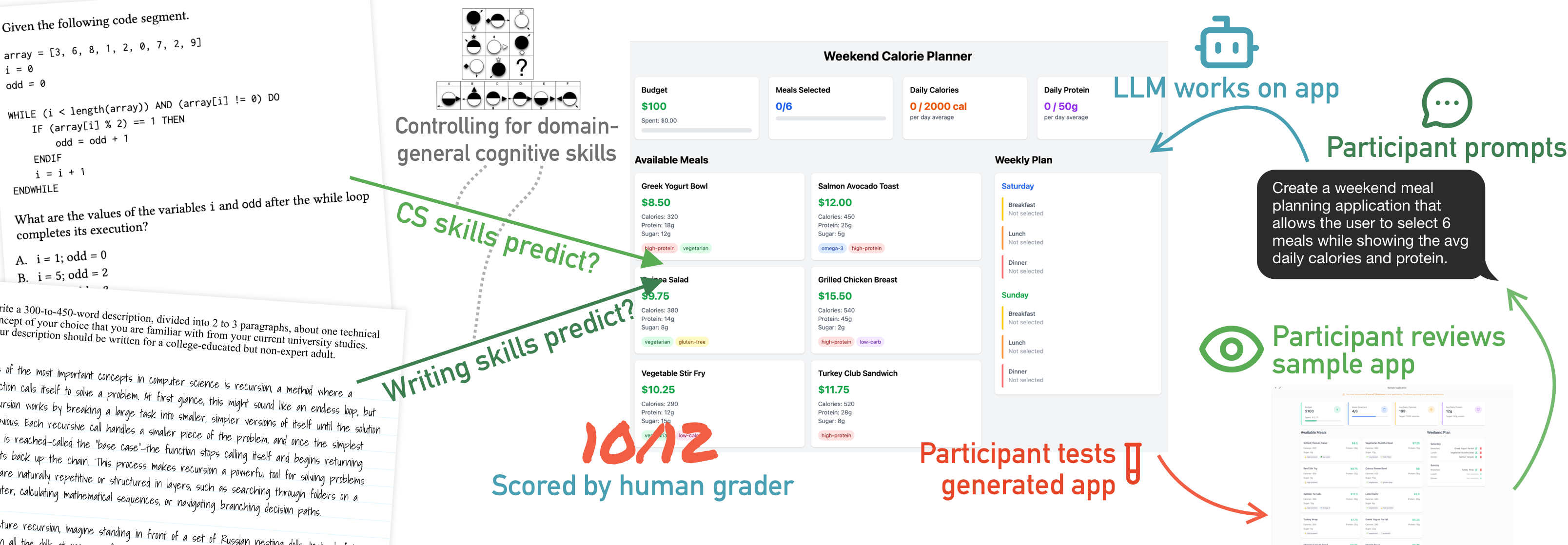}
 \caption{Our study investigates how writing skills and CS achievement relate to the ability to vibe code when controlling for domain-general cognitive skills. Participants reviewed sample applications, drafted prompts to an LLM-based agent, tested the resulting applications, and made further refinements. The final applications were then scored by a human grader.}
 \Description[Teaser figure giving an overview of the presented study]{The figure presents the study design connecting participant skills with their interactions with a large language model (LLM). On the left, participants' skills are assessed: a multiple-choice computer science question, and a short technical writing task. Handwritten notes ask whether "CS skills predict?" and "Writing skills predict?" outcomes. A small visual reasoning puzzle with the label "moderated by ICAR" is close to these notes and thin lines go to them. On the right, the workflow of participant interaction with the LLM is shown. A sample participant prompt requests a weekend meal planning application. The generated application in the center, titled "Weekend Calorie Planner". It is annotated with a grade of 10/12 and the note "scored by a human grader". The layout emphasizes the relationship between measured skills and subsequent performance in using and evaluating LLM-generated applications.}
 \label{fig:teaser}
\end{teaserfigure}

\author{Sverrir Thorgeirsson}
\authornote{Theo B. Weidmann and Sverrir Thorgeirsson are co-primary authors.}
\affiliation{
 \institution{ETH Zürich}
 \country{Switzerland}}
\email{sverrir.thorgeirsson@inf.ethz.ch}

\author{Theo B. Weidmann}
\authornotemark[1]
\affiliation{
 \institution{ETH Zürich}
 \country{Switzerland}}
\email{theo.weidmann@inf.ethz.ch}

\author{Zhendong Su}
\affiliation{
 \institution{ETH Zürich}
 \country{Switzerland}}
\email{zhendong.su@inf.ethz.ch}

\copyrightyear{2026}
\acmYear{2026}
\setcopyright{cc}
\setcctype{by}
\acmConference[CHI '26]{Proceedings of the 2026 CHI Conference on Human Factors in Computing Systems}{April 13--17, 2026}{Barcelona, Spain}
\acmBooktitle{Proceedings of the 2026 CHI Conference on Human Factors in Computing Systems (CHI '26), April 13--17, 2026, Barcelona, Spain}
\acmPrice{}
\acmDOI{10.1145/3772318.3791666}
\acmISBN{979-8-4007-2278-3/2026/04}

\maketitle

\section{Introduction}

Software development platforms with large language model (LLM) integration now let users describe programs in ordinary prose and obtain runnable systems with minimal code editing. As of today, such services are no longer niche; the AI-enhanced platform Replit reports having ``a global community of more than 40 million users''~\cite{replit2025funding}; Lovable, which is a similar service, states that its users have created more than 10 million projects~\cite{heim2025lovable}; and Bloomberg reports that the LLM-based IDE Cursor surpassed one million users in 2025~\cite{metz2025cursor}. The interaction style enabled by these and other platforms, dubbed \emph{vibe coding} by former OpenAI employee Andrej Karpathy~\cite{karpathy2025vibecoding}, allows one to rely on the behavior of the resulting program to iterate and improve the output program. As the capabilities of LLMs advance, this workflow is poised to become an even more common path for converting an idea into running code.

The shift toward language-first workflows raises questions about which skills matter most for future developers. For one, written communication skills may become increasingly important, in particular the ability to express intent unambiguously in writing. Unfortunately, this skill may currently be on the decline; for example, only 51\% of high school graduates in the United States reach the ACT’s English language benchmark for college readiness~\cite{ACT2024NationalGraduatingClassProfile} (a low not seen in decades~\cite{ACT2011ConditionCollegeCareerReadiness,ACT2016ConditionCollegeCareerReadiness,ACT2020ProfileReportNational}), and adult literacy scores in OECD countries have been on a downward trend for several years~\cite{OECD2024SurveyAdultSkillsUS}. To make sense of how this trend can affect current and future LLM-augmented software development, it would be helpful to know more precisely to what extent language skills, in particular written communication skills, relate to students’ effectiveness at vibe coding, and how the predictive ability of this construct compares to traditional measures of programming skills, namely computer science achievement. The answer has direct implications for tool and curriculum design, for instance, on the design of systems that teach users to write better prompts, and whether software-engineering curricula should place greater emphasis on written communication.

To address this question, we conducted a preregistered\footnote{The preregistration can be found at \url{https://aspredicted.org/wg3h-dx9m.pdf} with additional materials mentioned in the preregistration located at \url{https://doi.org/10.6084/m9.figshare.29509628}.}, cross-sectional study with 100 tertiary-level students recruited from several universities in the same metropolitan area (see Figure~\ref{fig:teaser}). Each student completed a battery of tests to measure CS achievement, domain-general reasoning, written-communication proficiency, and proficiency in vibe coding. Our vibe coding tasks were designed iteratively by eight experts who offered their input through a structured consensus-building process, and we asked students to solve them on a custom programming platform that we built after analyzing attributes of commercial AI-based programming platforms. In developing the platform and the tasks, we chose to concentrate on GUI-oriented vibe coding: participants used natural language to describe small, web-based graphical applications and iteratively refined them, without ever viewing or editing the underlying source code. We chose this focus because GUI-based app creation is a primary use case for contemporary vibe-coding platforms, and because GUI-based programming is one domain of vibe coding that provides observable behavior that can be evaluated on its own.

The primary aim of the experiment was to examine to what extent written communication skills and computer science achievement predict performance in vibe coding, and how much these predictions are moderated by general-reasoning skills. We also sought to understand how the predictive ability of student attributes is impacted by the nature of the vibe coding task in question; specifically, whether decontextualized tasks (tasks that intentionally lack context and therefore cannot be succinctly described but require detailed, feature-by-feature prompts) demand a greater level of written communication skills. Additionally, in line with the \emph{Standards for Educational and Psychological Testing}~\cite{american1985standards}, which emphasize collecting validity evidence based on response processes, we analyzed students' prompts to test whether prompt quality serves as a mechanism linking writing skills to vibe coding performance.

The contribution of our work is threefold:

\begin{enumerate}
    \item \textbf{Empirical answer to which skills drive GUI-oriented vibe coding performance}. 
    We provide a preregistered, cross-sectional study (N=100) that quantifies how written communication, computer-science achievement, and domain-general reasoning relate to performance in vibe coding, which we operationalized as programming through interacting with an LLM without concern for the underlying source code. We find that both CS achievement ($r = .39$) and writing skills ($r = .29$) significantly predict vibe coding performance, and that CS achievement remains a significant predictor after controlling for domain-general cognitive ability. In a joint model, CS achievement contributes roughly twice the unique variance of writing skills, though both add independent predictive value. The results have implications for both tool design and curriculum design (e.g., where to emphasize writing versus CS fundamentals).

    \item \textbf{A vetted assessment suite for GUI-oriented vibe-coding proficiency.} 
    We curate and document an expert-designed set of natural-language programming tasks produced via a structured consensus process with eight subject-matter experts. The set is accompanied by clear scoring criteria used in our study, and is intended to support replication, comparison across cohorts, and possibly instructional use.

    \item \textbf{A research platform for LLM-native software creation.} 
    We introduce a purpose-built environment that uses the core features of commercial AI programming tools while enabling controlled experimentation. The platform supports the iterative and behavior-guided workflow characteristics of vibe coding and is designed to support future empirical work on the construct.
    
\end{enumerate}

The paper is structured as follows. In Section 2, we offer some additional context for our research and describe related work. We will then describe how the tasks we deployed in the study were designed (Section 3), how the survey instruments were designed and selected (Section 4), and how our vibe coding platform was designed and how it functions (Section 5). We will then describe the design and results of our pilot study (Section 6), which informed aspects of our primary study, whose methodology and results are described in Sections 7 and 8. We then describe our exploratory results in Section 9 (including an analysis of the user-submitted prompts), discuss our findings in Section 10, and conclude the paper in Section 11.

\section{Background and Related Work}

\subsection{Vibe Coding}

The idea of using natural language to write computer programs predates LLM-guided programming, and has long been the subject of debate and criticism~\cite{halpern1966foundations,dijkstra1978foolishness}. Several lines of work have pursued this ambition from different angles and with varying success. The early programming language COBOL had the ambition to be ``an English language programming system designed for non-professional programmers''~\cite{sammet1961detailed}; Donald Knuth's ``literate programming'' paradigm~\cite{knuth1984} framed programs as ``works of literature,'' with code organized for human readers and embedded in explanatory prose~\cite{knuth1984}; and various strands of research under the name of ``natural-language programming'' have explored translating constrained English into executable code, often via controlled subsets of English to avoid ambiguity~\cite{kuhn2014survey}. However, only with the advent of LLMs has it become feasible to map conversational (and possibly underspecified) instructions to code.

LLM-augmented programming can take many rich forms, and not all implementations of the paradigm expect the user to communicate entirely in prose. For example, Liu et al.'s~\cite{liu2023what} workflow for spreadsheet programming converts a user's prose to code and then reflects it back as standardized, editable natural-language steps, letting programmers switch easily between code or prose, depending on the abstraction level they prefer. In the same vein, a recent paper on a system called \emph{DirectGPT}~\cite{masson2024directgpt} introduced the idea of allowing users to modify parts of the generated artifact both with natural-language prompting and direct manipulation on the visual output of the program, allowing the programmer to choose whether to focus on prose or the system output. However, many LLM-augmented development workflows support coding more as a chatbot interaction. Jiang et al.'s 2022 work~\cite{Jiang2022}, identified in a recent scoping review as the earliest academic paper on an LLM-based coding tool~\cite{lau2025designspace}, found that users effectively treat natural-language prompts as a new programming notation, developing ad hoc ``syntax'' and debugging strategies for their prose instructions when working with a code-generating LLM. Although it is not clear whether this result still holds (other findings from that era have been conjectured to be ``not as relevant anymore'' because LLM output quality has significantly increased~\cite{berke2023}), recent work continues to show that LLM-based programming tools can substantially reshape the way humans approach software development. For instance, Mozannar's 2024 study found that using LLM-based tools for programming, such as GitHub Copilot, can ``significantly change user behavior''~\cite{Mozannar2024}, and that developers differ in how they choose to use such tools: some write natural-language descriptions of individual functions, whereas others rely on more technical notation such as docstrings and function signatures. Taken together, these results suggest variation not only in the design of LLM-augmented programming tools, but also in how developers use them in practice.

More specifically, the interaction style in which the programmer does not interact with the code, instead relying on the LLM to generate a working program, was dubbed \emph{vibe coding}~\cite{karpathy2025vibecoding} by OpenAI co-founder Andrej Karpathy in February 2025. Karpathy's formulation involves that one will ``forget that the code even exists,'' and as such, vibe coding can be considered a variant of LLM-augmented programming that relies exclusively on iteratively writing prompts and evaluating program behavior. Consequently, it means that the act of writing turns into a primary programming skill: developers must translate design intent into precise, revisable prose instructions for the model. The viability of vibe coding means that building software is now possible for those without the technical skills needed to understand code, as several platforms that support vibe coding have made explicit; for example, Replit's blog post on vibe coding describes it as ``a paradigm shift in app development'' under which one no longer needs to ``learn programming languages and syntax'' or ``understand complex technical concepts''~\cite{palmer_what_2025}, and Replit's CEO has claimed that 75\% of its users never write a single line of code~\cite{masad_vibe_coding_2025}. 

As a new term, the definition of vibe coding is still in flux and susceptible to semantic dilution; some define the term broadly as any form of AI-guided programming~\cite{collins:vibe-coding}, while others insist that such definitions dilute the term and that vibe coding refers exclusively to ``building software with an LLM without reviewing the code it writes''~\cite{willison_vibe_2025}. The Google Cloud documentation~\cite{googlecloud_what_is_vibe_coding} posits that vibe coding comes in two variants. One is the ``pure'' vibe coding version, which is consistent with Karpathy's definition, while the other version means that the user understands the code generated by the AI and can refine it. In our study, we explicitly target this ``pure'' or ``no-code'' variant of vibe coding, as it provides a well-defined construct to measure: vibe-coding proficiency as the ability to iteratively specify, refine, and debug program behavior through text prompts and observed behavior alone. A more permissive workflow that exposes the source code would risk confounding this construct, as some participants might choose to edit the generated code directly through the LLM interface, effectively reverting to traditional programming.

Vibe coding has many potential use cases and application domains. For instance, using vibe coding to build an application with a GUI remains a primary use case for LLM-based commercial systems such as Replit and Lovable, which provide integrated support for the front-end web stack (HTML, CSS, and JavaScript). This setting can exploit the verification-generation asymmetry of vibe coding: GUI-based applications have observable behavior or design that can be easy for users to evaluate iteratively, but can be hard or time-consuming for a human to generate. Vibe coding is also used for specifying data transformations and data analysis, for example in a similar way that Liu~\cite{liu2023what} describes, with the use of designing formulas in natural language for spreadsheets, or even for applications where the primary object of interest is a physical artifact rather than source code; previous work has shown that LLM-augmented programming can be used successfully for programming robots~\cite{Karli2024}. In our work, since vibe coding is widely used for visual applications, as seen by the large number of platforms focused on this use case, we chose to scope our evaluation to GUI-oriented vibe coding. One of our tasks also involved algorithms, so our task suite captures a limited subset of more algorithmic-centric vibe coding; we return to this boundary condition in the discussion.

\subsection{Skills Required for Vibe Coding}

The role of prose writing for vibe coding makes it a salient research question to what extent writing skills play a role in the quality of the applications that are generated. According to the National Assessment Governing Board that serves the U.S. Department of Education, writing is a ``complex, multifaceted, and purposeful act of communication'' that can be accomplished in many environments and with many different tools~\cite{NAGB2010WritingFramework}. Writing skills are measured by various criteria, for example in how it meets ``audience, purpose and genre conventions'' which can be assessed in relation to content criteria such as organization and structure, reasoning, evidence, mechanics, syntax and surface conventions used in the writing~\cite{AACU2009WrittenCommunicationVALUE}, but also more simply in how successfully one is able to ``explain in order to expand the reader's understanding''~\cite{NAGB2010WritingFramework}. In our work, we are particularly interested in the latter, which includes the ability to write reasoned and logically structured content, as success in vibe coding is likely to be at least partially determined by the ability to explain intent to a non-human collaborator that does not share the writer's tacit context. For human-to-human software requirement engineering, the importance of this ability is well-understood; the IEEE International Standard mandates that requirements should be ``stated simply'' and be ``easy to understand''~\cite{ieeestandards}, and the NASA System Engineering Handbook~\cite{nasa} appendix on ``How to Write a Good Requirement'' offers specific language guidance, for example, to avoid indefinite pronouns that could lead to misinterpretation.

Studies on human-generated prompt quality to LLMs suggest that many non-experts lack robust strategies for the kind of technical writing required by this new workflow, exploring prompts ``opportunistically, not systematically'' and importing misleading assumptions from everyday conversation~\cite{zamfirescu2023johnny}. Feng et al.'s study of the CoPrompt system shows that even experienced programmers struggle with prompt clarity and level of detail when using natural language to program: participants often wrote vague, ``unorganized'' prompts whose relationship to the resulting code was hard to reconstruct, and needed tooling to make the prompts readable and reusable~\cite{feng2024coprompt}. Furthermore, some studies have analyzed explicitly how the quality of text impacts performance on LLM-guided programming, though not, to the best of our knowledge, via direct, participant-level assessments of writing proficiency in a cross-sectional design. Lucchetti et al.~\cite{Lucchetti2025} analyzed properties of novice-written prompts and found that information content, not technical vocabulary, best predicts success; substituting imprecise terms with more precise terms from existing prompts had little effect on how the LLM performed, but miswordings that affected the semantics of a prompt (e.g. ``print'' instead of ``return'') did affect pass rates. They also show that rewording without adding information rarely helps and is common in failure cycles. A recent work found that LLM performance on code tasks degrades significantly by introducing syntax and grammar errors, but also structural variations in human language, such as the use of interrogative instead of declarative statements~\cite{Chen2025NLPerturbator}. 

The impact of computer programming or CS skills on LLM-assisted programming performance has been studied directly and indirectly in several other research papers. A 2025 work studied the performance of those with and without programming experience on communicating input-output transformation problems in natural language, both to an LLM and to another human~\cite{Pickering2025}. Although the style of communication (e.g., a generous use of examples) was a more important attribute than expertise, the study also found that those with a programming background had a modest performance advantage. Similarly, a 2024 study~\cite{Nguyen2024} found that exposure to CS beyond an introductory course predicted higher success on a specific style of programming tasks. Participants were given input-output pairs that corresponded to CS1-style programming tasks and asked to write natural language prompts to an LLM to generate code. Similar to our study, participants could not work with the generated code directly, as the objective was to ``[focus] on natural-language-to-code interactions.'' Unlike our study, the LLM system relied on a single-turn, stateless prompting, meaning that while participants could make unlimited attempts, they could not build on prior context. This design differs from typical LLM workflows, but may have introduced a tighter experimental control.

A correlational research study from 2023~\cite{Jeuring2023} found that performance on a computational thinking test~\cite{wiebe2019development} (developed for middle school students but administered to adults) predicted performance on LLM-guided programming. The correlation was strong and significant, but had limited precision due to the low sample size ($N=19$). As in our study, the participants were presented a visual task (a simple-grid based game) and were then asked to recreate it iteratively with the assistance of an LLM, and the quality of the program was measured with a functionality criteria. However, unlike our study, the authors note that the limitations of the online programming environment made it impossible to test a user interface; instead, the participants had to assess the functionality of the resulting program via text.

In general, language skills may have some connection to programming performance. A 2014 study~\cite{siegmund2014understandingA} used functional magnetic resonance imaging (fMRI) to show that brain regions that are related to language processing are activated during programming, although this result was contradicted by later findings~\cite{liu2020computer,ivanova2020comprehension}. English language skills have been found to correlate with software engineering achievement~\cite{Alrasheed2021}, and according to a 2020 study, adult learners’ language-learning aptitude was a greater predictor of performance in Python programming than mathematics skills~\cite{prat2020relating}, which is consistent with older findings that verbal achievement is a greater predictor of CS grades than math achievement~\cite{leeper1982predicting}. However, we are not aware of existing studies on the relationship between writing ability alone and CS achievement, LLM-augmented or otherwise. 

It is likely that domain-general cognitive skills may influence both writing and CS achievement. According to a 2023 study on 282 undergraduate students, certain types of cognitive skills had a significant, positive correlation with performance on a CS achievement test (namely the SCS1, which we also used in our work)~\cite{Graafsma2023}. Furthermore, writing skills and cognitive skills have been found to be correlated; a study on 550 children found a large, positive relationship (Pearson's $r$ = 0.77) between intelligence test results (WISC-IV FSIQ~\cite{wechsler1991wechsler}) and writing test results (the WIAT-II's writing component~\cite{wechsler2001wiat}). While these results come from different populations, they suggest that general cognitive skills could be a shared factor behind all three outcomes. To address this for our study, we included a measure of cognitive skills as a control variable so that any links we observe between CS achievement, writing, and vibe coding skills are less likely to be explained simply by domain-general cognitive skills. In our work, we include cognitive skills only as a statistical control to reduce confounding, not as a statement about its instructional importance or a recommendation for educators to consider such scores in their instruction.

\section{Task Design}
\label{section:tasks}

As a measurable construct, vibe coding lacks a universal definition, task designs, and performance criteria. In our study, we chose to evaluate the construct by (a) granting the participants access to an LLM-based agent that writes code iteratively from human task specifications, and (b) asking the participants to solve two types of tasks: replication and feature extension of existing programs that students could explore at will. This choice of task types was motivated by both practical and measurement considerations. For one, providing a textual specification would invite students to copy that text directly into the model or treat it as the definitive prompt, which would shift the construct to the near-trivial (or at least irrelevant) ability to copy or paraphrase text. Instead, our design requires the participants to formulate the target themselves, evaluating their ability to articulate a system design in such a way that an LLM agent can implement it. Broadly speaking, the ability to infer behavior from an existing system has many established use cases in software development and security~\cite{Klimek2011}, and has precedents in computing education, for example in reverse-engineering games~\cite{aycock2015applied}.

To acquire the vibe coding tasks that the participants were meant to replicate or extend, we used a Delphi-style technique~\cite{linstone1975delphi}, namely expert elicitation using a two-stage structured process. Our goal with this process was to acquire two of the three tasks that we would use for our study. After securing ethics approval, we recruited eight participants for a task generation expert panel; all panelists were recruited based on their substantial expertise with LLMs and vibe coding. No compensation was offered. In addition to their familiarity with the domain, the experts were selected for their geographic and occupational diversity; the panelists were based in four countries and included three professors in the fields of CS education and human-computer interaction, one secondary school CS instructor, one postdoctoral researcher, one doctoral student, one master's student, and one front-end developer. Four of the panelists were women and four were men. To preserve independence, the authors of this paper did not serve on the panel. 

In the first stage of the process, each panelist was asked to offer a full specification of a task that met the following criteria:

\begin{enumerate}
\item \textbf{Interdisciplinarity}: The topic of the task should resonate beyond computer science, engaging at least one other domain (e.g., linguistics, civics, or art).
\item \textbf{Real-world use case}: The task should be grounded in a plausible scenario that reflects a genuine need or common challenge encountered outside academic or very specialized contexts.
\item \textbf{GUI-first implementation}: The resulting program should be used primarily through a graphical user interface (GUI); external APIs are acceptable, but no server-side programming should be required.
\item \textbf{Non-trivial complexity}: The task should involve multiple components and decision points, so it is not trivial with LLM assistance, yet avoid advanced algorithms and keep mathematics at or below the secondary-school level.
\item \textbf{Time-feasible}: Participants should be able to implement an end-to-end version within roughly 10--15 minutes using iterative prompting.
\item \textbf{Clear specification}: A typical tertiary-level student should be able to grasp the app's purpose and required features quickly, minimizing time spent interpreting the brief.
\item \textbf{Culture-fair}: The task should not assume particular cultural experiences or local knowledge, reducing construct-irrelevant variance and supporting fair comparison across participants.
\end{enumerate}

We chose the first two criteria to keep the tasks authentic; in Schriebl et al.'s multidimensional model of authentic learning in science education~\cite{Schriebl2023}, ``real-world authenticity'' is strengthened by content contextualization, especially societal relevance, connection to daily life, and content-related interdisciplinarity. The third criterion, that the tasks should possess a GUI, was important for several reasons; first, we find that a core aspect of vibe coding is the iterative shaping of how a program feels in use, which is inferred from its state and responsiveness. To achieve this, a visible interface is more helpful than to infer the state from code or console output. Second, guidance on authentic tasks also emphasize a visual product that students can share, a feature that is naturally supported by applications with a GUI, and third, we believe that vibe coding is often used for such applications in practice (for example on existing ``vibe coding platforms''; see Section~\ref{section:platforms}). The remaining criteria were designed to help us run a fair, timed study, keeping the focus on what participants do with LLM support, and reducing construct-irrelevant variance, such as task misreading.

\label{task-design}

\newcommand{\colorrating}[2]{
  \ifdim #1 pt >4.5pt \cellcolor{green!25}{#1 (#2)}
  \else\ifdim #1 pt >4.0pt \cellcolor{green!10}{#1 (#2)}
  \else\ifdim #1 pt >3.5pt \cellcolor{yellow!20}{#1 (#2)}
  \else \cellcolor{red!15}{#1 (#2)}
  \fi\fi\fi
}
\begin{table*}[t!]
\centering
\renewcommand{\arraystretch}{1.2}
\begin{tabular}{l|c|c|c|c|c}
\toprule
\textbf{Task} & \textbf{Interdisc.} & \textbf{Authenticity} & \textbf{Challenge} & \textbf{Purpose clarity} & \textbf{Culture fairness} \\
\midrule
\rowcolor{blue!15}
Task 1: Exam scheduler      & \colorrating{4.3}{1.0} & \colorrating{4.4}{1.0} & \colorrating{4.7}{0.5} & \colorrating{4.3}{1.0} & \colorrating{4.6}{1.0} \\
Task 2: Currency calculator & \colorrating{4.3}{0.5} & \colorrating{3.9}{2.0} & \colorrating{3.7}{2.0} & \colorrating{4.7}{0.5} & \colorrating{4.4}{1.0} \\
\rowcolor{blue!15}
Task 3: Meal planner        & \colorrating{4.6}{0.5} & \colorrating{4.1}{1.0} & \colorrating{4.1}{1.5} & \colorrating{4.6}{1.0} & \colorrating{4.6}{1.0} \\
Task 4: Meeting organizer   & \colorrating{4.3}{0.5} & \colorrating{4.4}{1.0} & \colorrating{3.7}{1.5} & \colorrating{4.0}{1.5} & \colorrating{4.6}{1.0} \\
Task 5: Isometric room builder & \colorrating{4.1}{0.5} & \colorrating{3.3}{1.0} & \colorrating{3.7}{1.5} & \colorrating{3.9}{0.0} & \colorrating{4.7}{0.5} \\
Task 6: Eye tracker app     & \colorrating{3.7}{2.5} & \colorrating{3.7}{2.0} & \colorrating{3.7}{2.5} & \colorrating{2.6}{2.0} & \colorrating{3.7}{2.0} \\
Task 7: Celebrity ranker    & \colorrating{3.7}{1.0} & \colorrating{3.3}{1.0} & \colorrating{3.6}{1.0} & \colorrating{4.6}{1.0} & \colorrating{4.4}{1.0} \\
\rowcolor{blue!15}
Task 8: Course platform & \colorrating{4.4}{1.0} & \colorrating{4.6}{1.0} & \colorrating{4.3}{1.0} & \colorrating{4.7}{0.5} & \colorrating{4.9}{0.0} \\
\bottomrule
\end{tabular}
\vspace{1em}
\caption{The mean panelist ratings and interquartile ranges (in parentheses) of each proposed task according to our five criteria, namely interdisciplinarity, real-world use case (authenticity), non-trivial level of complexity (challenge), purpose clarity, and culture fairness. The tasks that were selected for the study (i.e., an exam scheduler, meal planner, and a course registration platform) are highlighted.}
\label{tab:delphi2}
\end{table*}

In the second round of the process, each panelist was asked to rate each other's tasks anonymously on a five-point Likert scale using five scales from the earlier round. We chose to omit the third scale for this round as all provided tasks met this criterion, and the fourth and fifth scale were collapsed into one. After the second round was complete, we found three tasks that scored high on each scale (median rating $\ge 4$ on all five scales) and whose ratings had a high consensus, i.e., the interquartile range (IQR) was $\leq$ 1 for at least 80\% of the scales, making a third round unnecessary and likely to impose an unnecessary burden on the panelists. The average ratings can be seen in Table~\ref{tab:delphi2}. Two of these tasks (Task 1 and Task 8) had a similar description, so we chose to combine them into one. Our panelists also offered comments on how each task could be improved, some of which we used to refine the tasks; for example, the meal planner application was missing an ability for students to select specific meals.

Lastly, to complement the two authentic tasks selected via the Delphi-like process, we designed a third decontextualized task that was intentionally non-authentic. The aim was to stress the construct of vibe coding as expressive specification: we wanted to determine if the participants would be more successful at implementing tasks with vibe coding without relying on task labels that carry strong priors for an LLM, such as a ``meal planner'' or a ``scheduler.'' To that end, we designed an additional decontextualized task in the form of a small GUI toy application that was designed to have obvious behavior, but the labeling of the user interface was intentionally opaque as to not reveal a specific purpose.

Submissions were evaluated for behavioral fidelity using rubrics that we defined before the study took place. Participants were instructed that imitating the style and design of the base applications, such as the colors and dimensions of their components, was not necessary.

\section{Survey Instruments}

\subsection{Written Communication Skills}

Identifying a suitable instrument to measure adults' writing proficiency is challenging for two reasons: construct fit and deployment constraints. First, the literature documents a persistent gap in adult-focused, standardized measures~\cite{mcnair2013forgotten}; many widely used batteries stop providing reference data beyond very early adulthood and/or use developmentally inappropriate prompts (such as WIAT-III's ``what is your favorite game?''~\cite{hebert2016examining,mcnair2013forgotten}). Second, even for nearly acceptable instruments, practical considerations made them unsuitable for our study design; many instruments are proprietary and restricted to specific purchasers, or have licensing, privacy, and access issues, such as requiring participants to use an external website where we do not have control over user data.

For this reason, we enlisted two instructors in written communication from the \emph{Language Center of UZH and ETH Zurich} to design a task-based assessment and an analytic rubric for adult writers based on assessments used in their course. The instructors were volunteers, not a part of the author team of this paper, and were not compensated for their help. The rubric they designed can be found in Figure~\ref{fig:writing-rubric}. The essay prompt is:

\begin{quotation}
\small\itshape
Write a 300-to-450-word description, divided into 2 to 3 paragraphs, about one technical concept of your choice that you are familiar with from your current university studies. Your description should be written for a college-educated but non-expert adult.
\end{quotation}

To understand the construct coverage of this assessment, we consulted with staff responsible for AAC\&U’s VALUE initiative~\cite{AacuValueRubrics,mcconnell2019we} about the Written Communication VALUE rubric~\cite{AACU2009WrittenCommunicationVALUE}, which has been used by many universities in the United States. While the VALUE rubric was not a good fit for our needs as it is designed for research-supported writing and institution-level assessment rather than grading, the criteria of the rubric (context and purpose, content development, genre/disciplinary conventions, sources and evidence, and control of syntax and mechanics) is close to the one we used. This conceptual overlap, together with VALUE's explicit expectation that campuses translate and adapt rubric language to local contexts, lends some support for the content validity of our locally developed instrument.

To grade participant-written essays according to the rubric, two independent graders rated each essay blindly and independently; one professional editor and technical writer, and one graduate assistant. We held two calibration meetings for the graders to clarify questions on the rubric, such as how to grade single-paragraph essays. Per our preregistration, we targeted inter-rater reliability of at least $ICC(2,2) \ge 0.75$. When the two ratings differed by more than ten points, a third blinded rater (a professor in linguistics, also independent) scored the essay, and the final score was the mean of the two closest ratings (otherwise, the mean of the first two was used). Our pilot study helped us settle the final wording of the prompt and the amount of time allotted, which we then held constant for all participants.

\begin{figure}[t!]
  \centering
  \begin{minipage}{0.97\columnwidth}
    \begin{framed}
        \begin{enumerate}[leftmargin=1.2em, labelsep=0.4em, itemsep=0.5em]
        \item \textbf{Overall effectiveness for the target readership}
          \begin{enumerate}[leftmargin=1.2em, labelsep=0.4em, itemsep=0.35pt, topsep=0.2em]
            \item The scope of the topic is appropriate for the length and timing of the exercise.
            \item The topic is explained with an appropriate level of complexity (i.e., not over-simplified, but not too detailed).
            \item New or difficult terminology is adequately explained.
            \item The average target reader would understand this text.
          \end{enumerate}
        \item \textbf{Paragraphs}
          \begin{enumerate}[leftmargin=1.2em, labelsep=0.4em, itemsep=0.35pt, topsep=0.2em]
            \item The paragraphs address topics in a logical order.
            \item Each paragraph has either a topic sentence or another clear organizational structure.
            \item Each paragraph addresses a unified, cohesive topic.
            \item There are visible and logical transitions between paragraphs.
          \end{enumerate}
        \item \textbf{Sentences}
          \begin{enumerate}[leftmargin=1.2em, labelsep=0.4em, itemsep=0.35pt, topsep=0.2em]
            \item Sentences are clear and straightforward, with subjects and verbs that can be easily identified.
            \item The majority of sentences use active verbs that appear in the first half of the sentence.
            \item The writer varies sentence structures, including both complex and simple sentences.
            \item The writer varies sentence length.
          \end{enumerate}
        \item \textbf{Flow between sentences}
          \begin{enumerate}[leftmargin=1.2em, labelsep=0.4em, itemsep=0.35pt, topsep=0.2em]
            \item Topics follow a logical order from sentence to sentence within each paragraph.
            \item The links between sentences are clearly identifiable (i.e., no unclear pronouns, unclear continuations, etc.).
            \item The beginning of most sentences has a clear link to the previous material.
            \item A range of cohesive devices (e.g., repetition, substitution, anaphoric nouns, etc.) is used flexibly and accurately.
          \end{enumerate}
        \item \textbf{Language-level concerns}
          \begin{enumerate}[leftmargin=1.2em, labelsep=0.4em, itemsep=0.35pt, topsep=0.2em]
            \item The vocabulary is appropriate (in terms of complexity) for the target audience.
            \item The vocabulary and grammar are accurate and precise enough for the reader to easily understand the text. Errors are minor and rare and have minimal impact on clarity.
            \item The vocabulary and grammar are varied.
            \item A wide range of lexical and grammatical structures is used accurately and flexibly to convey clear and precise meanings.
          \end{enumerate}
      \end{enumerate}
    \end{framed}
  \end{minipage}
  \caption{The grading rubric of written communication skills that we deployed in the study after expert consultation. Each sub-item is graded on a scale from 1 to 5, resulting in a maximum score of 20 for each category, and a maximum score of 100 in total.}
  \label{fig:writing-rubric}
\end{figure}

We note that we use the rubric to index a domain-general proficiency in written communication. Individual criteria, such as paragraph organization and cross-sentence cohesion, are treated as reflective indicators of that proficiency. We do not claim that any specific feature (e.g., use of anaphoric nouns) is itself beneficial for LLM interaction; rather, we test whether the underlying ability to structure and convey technical information in prose predicts performance on LLM-mediated programming tasks.

\subsection{Computer Science Achievement}

To evaluate CS achievement, we conducted a search of validated instruments used in computing education. We sought an established instrument that made minimal assumptions about participants' prior programming experience, i.e., which specific programming languages they had been instructed in. Ultimately, we chose to measure the construct using a 12-item subset of the SCS1, a validated, language-independent (pseudocode-based) assessment of CS knowledge developed by Parker et al.~\cite{parker2016replication}. Since its inception, SCS1 has become an important benchmark for measuring CS learning; a 2021 review identified 44 studies that have utilized this assessment in varied research contexts~\cite{parker2021uses}.

The 12-item subset that we used was the same one that Parker et al.~\cite{parker2018socioeconomic} used in another correlational, cross-sectional study to measure CS achievement. In choosing this subset, we were also motivated by Parker et al.'s rationale, namely to maintain the content coverage of the full 27-item test by including four items each on definitional knowledge, code tracing, and code completion, and to preserve psychometric quality by selecting items according to their previously estimated item-response-theory difficulty and discrimination values. Furthermore, we believed that administering the full SCS1 in addition to the other tasks and surveys would induce too much testing fatigue, as the SCS1 is meant to take an hour to complete~\cite{parker2021uses}. Hence, for the subset, we chose to impose a 25-minute time limit, which is proportional to its length. To ensure that, in doing so, we would avoid construct-irrelevant speed, we gathered validity evidence of the timed administration of the assessment in our pilot study.

\begin{figure*}[t!]
\centering
\setlength{\FrameSep}{6pt}   
\setlength{\FrameRule}{0.4pt}
\begin{framed}
\begin{minipage}[t]{0.45\textwidth}
  \footnotesize
  \textbf{Question 12 from ICAR16}
  \par\vspace{0.3\baselineskip}
  Please indicate which is the best way to complete the figure below.
  \par\vspace{0.5\baselineskip}
  \centering
  \includegraphics[width=0.95\linewidth]{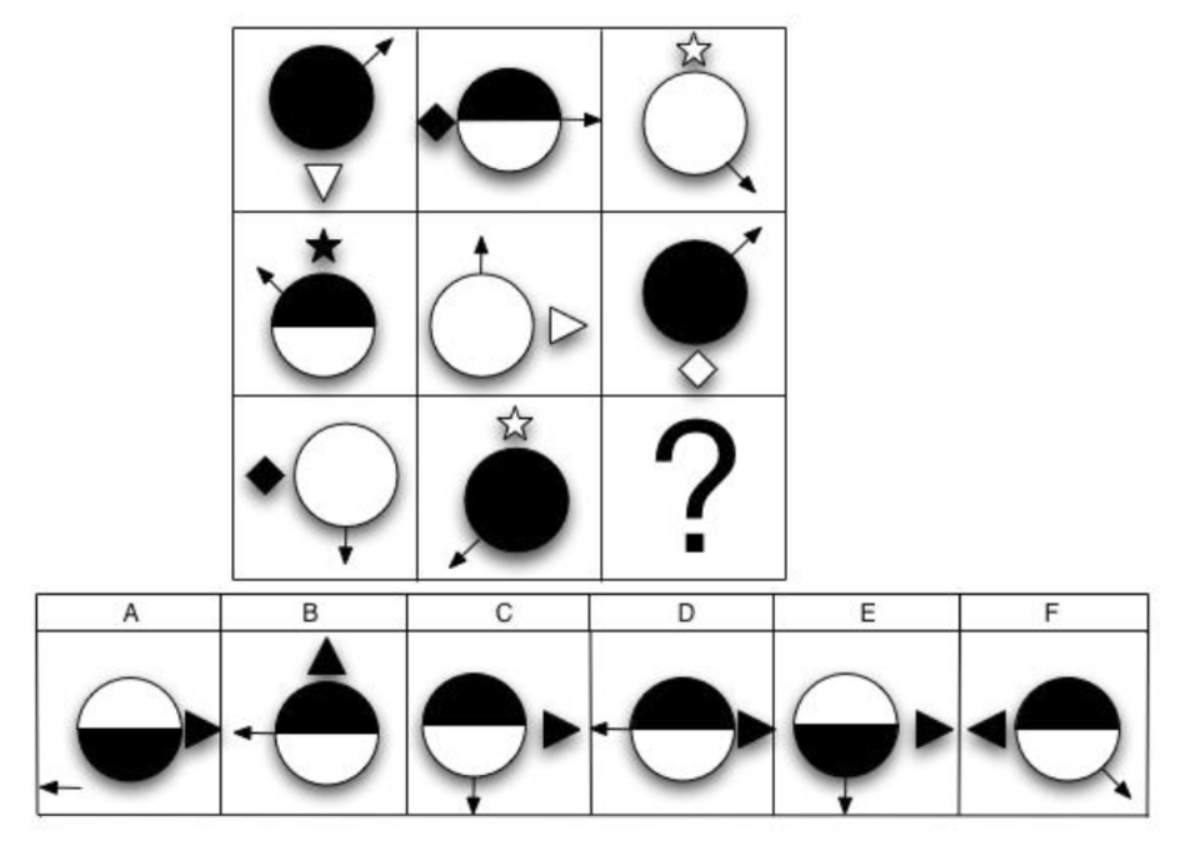} \\
  \footnotesize
  Answer options: (1) A (2) B (3) C (4) D (5) E (6) F

  \medskip
\end{minipage}\hfill
\begin{minipage}[t]{0.45\textwidth}
  \footnotesize
  \raggedright
  \textbf{Question 9 from SCS1}
  \par\vspace{0.3\baselineskip}
  Given the following code segment.
    \vspace{0.5\baselineskip}

\begin{lstlisting}[basicstyle=\ttfamily\scriptsize, % smaller code
                   columns=fullflexible,
                   keepspaces=true,
                   showstringspaces=false,
                   breaklines=true,
                   tabsize=4,
                   language=Python,
                   keywordstyle={}]
array = [3, 6, 8, 1, 2, 0, 7, 2, 9]
i = 0
odd = 0

WHILE (i < length(array)) AND (array[i] != 0) DO
    IF (array[i] % 2) == 1 THEN
        odd = odd + 1
    ENDIF
    i = i + 1
ENDWHILE
\end{lstlisting}

  \vspace{0.6\baselineskip}

  What are the values of the variables \texttt{i} and \texttt{odd} after the while loop completes its execution?

  \vspace{0.5\baselineskip}

\begin{enumerate}[label=\Alph*. , leftmargin=1.4em, itemsep=0.2em, topsep=0.2em]
  \item i = 1; odd = 0
  \item i = 5; odd = 2
  \item i = 5; odd = 3
  \item i = 8; odd = 4
  \item i = 8; odd = 5
\end{enumerate}

\end{minipage}
\end{framed}
\caption{Items from assessment instruments ICAR16 (left) and SCS1 (right) that were provided as examples in the ICAR guidelines~\cite{ICAR2014} and by Parker et al.~\cite{parker2016replication}, respectively.}
\label{fig:assessment-side-by-side}
\Description[Items from ICAR16 and SCS1]{The left image shows a matrix-reasoning question in which the test-taker is asked to identify which of a list of shapes is missing from a matrix. There are 6 available answers.}
\end{figure*}

\subsection{General Reasoning Skills}

To measure domain-general cognitive skills, we used an assessment instrument called the International Cognitive Ability Resource (ICAR) that was developed as an open alternative to proprietary tests~\cite{condon2014international}. The test consists of multiple-choice questions that involve verbal reasoning, matrix reasoning, letter series, and three-dimensional rotation. Compared to licensed options that also measure cognitive skills, ICAR is attractive to researchers as it requires no licensing fees, and its public availability (upon request) increases research transparency and reproducibility.

ICAR consists of 60 questions, but its developers also constructed a 16-question subset of the measure called the ICAR16, which contains four questions of each type~\cite{condon2014international}. Empirical studies support the ICAR16 as a valid indicator of domain-general reasoning; the test shows strong convergence with comprehensive cognitive batteries, with one cross-sectional study finding a large correlation ($r=0.81$) between scores on ICAR16 and the proprietary Wechsler Adult Intelligence Scale (WAIS-IV)~\cite{young2020}. The instrument has also been found to be age and sex-invariant~\cite{young2019}. ICAR16 has been deployed in a number of research contexts; for instance, it has been deployed in genetics~\cite{Willoughby2021Genetic} and mental health~\cite{keidel2024} research.

The ICAR16 should be administered with no more than a 16-minute time limit in studies involving young adults, and some researchers use as little as 10 minutes~\cite{dworak2021using}. To find a suitable timer, the ICAR designers recommend piloting the test with a ``handful of participants in your target population''~\cite{ICAR2014}, which is advice that we followed for our study (see Section~\ref{section:pilot}). After our pilot, we chose a time limit of 12 minutes in our main study. The order of the 16 items was randomized as in Condon and Revelle's initial deployment~\cite{condon2014international}. 

\section{Vibe Coding Platform}

We developed a dedicated vibe-coding platform to guarantee reproducible access, satisfy privacy requirements, and to give us greater control over how to instantiate vibe coding. Reproducibility concerns included version drift, evolving interfaces, commercial paywalls, and the risk that services change or disappear during a study. We also wished to meet our ethics committee's requirements for robust data protection; third-party platforms provide more limited transparency and rely on varying data-handling policies. Importantly, we had to ensure that the provider of the AI model integrated into our tool does not retain participant inputs and that the model is not trained on participants' prompts.

\subsection{Survey of Prominent Vibe-Coding Tools}
\label{section:platforms}

To ensure that our research platform was informed by the interaction flow and visual layout of existing tools, we first analyzed four commonly cited platforms associated with vibe coding. In addition to direct use of LLM chatbots such as ChatGPT, prominent tools include Replit, Cursor, Lovable, and Bolt \cite{sarkar2025vibe, geng2025exploring}, which have also been highlighted in articles by major news outlets such as the New York Times and the Wall Street Journal~\cite{roose2025not, sun2025your}. To compare them systematically, we asked each platform’s AI agent to implement the same toy application of a simple password manager. This task was chosen because it requires both interface design and code generation, yet remains small enough to complete within a single session. While using each tool, we performed screen recordings and noted how the system supported iteration, debugging, and user control.

Across the four platforms we examined, the user interfaces followed a highly similar structure. Each relied on a chat-based interaction with the AI agent, coupled with a live view of the application under development. In Replit, Lovable, and Bolt, the chat window was placed on the left, with the application preview occupying the remainder of the screen. Cursor differed slightly: marketed as an ``AI Code Editor''~\cite{cursor2025homepage} and built on top of Visual Studio Code~\cite{cursor2025vscode}, it displayed the code editor on the left and the chat interface on the right. The four systems made it possible to view the source code if the user so desired, and, depending on the subscription plan, some of them allowed one to also edit the underlying code. We intentionally diverged from commercial tools here, as exposing the source code would have introduced an additional interaction mode and risked confounding our construct of ``pure'' vibe coding.

\subsection{Study Platform Design}

\begin{figure*}[t!]
    \centering
    \includegraphics[width=\linewidth]{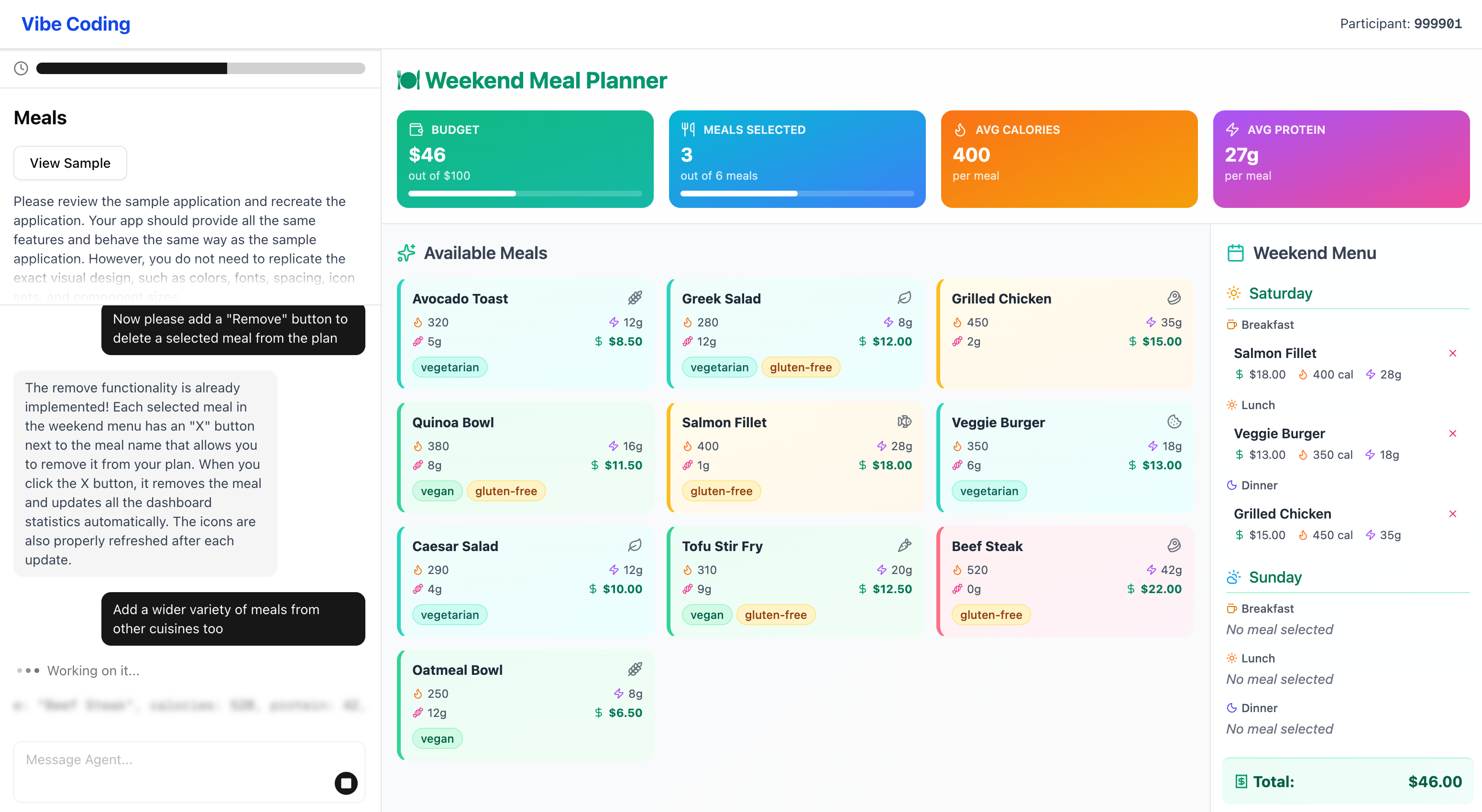}
    \caption{User interface of the tool we used in the study. The interface follows the vibe coding paradigm by not exposing any code. It features a chat box on the left, consistent with other tools, and an app preview on the right. The left sidebar also includes the task description and a timer showing the remaining time for participants (hovering over the bar shows the exact time remaining). The interaction seen in the screenshot was recreated based on prompts from participants during the study.}
    \Description[Screenshot of the vibe coding platform built for the study]{The screenshot shows the user interface of the vibe coding platform. On the left, there is a chat box where participants interact with the system, along with the task description at the top and a progress bar timer. At the bottom left a very blurred code live stream indicates progress by the AI agent. The right side displays the generated application titled "Weekend Meal Planner." At the top, four summary boxes track budget, meals selected, average calories, and average protein. Below, a grid of available meals shows items such as Avocado Toast, Greek Salad, and Beef Steak, each with calories, protein, sugar, price, and dietary tags. On the far right, a weekend menu panel lists Saturday and Sunday with empty slots for breakfast, lunch, and dinner, as well as a total cost field at the bottom.}
    \label{fig:ourtool}
\end{figure*}

In contrast to the prominent platforms we evaluated, our own platform deliberately conceals the code entirely, consistent with the original definition of vibe coding~\cite{karpathy2025vibecoding}. As shown in Figure~\ref{fig:ourtool}, the interface retains the chat window on the left and the app preview on the right, following the conventions of existing tools. To support our study design, we extended the left-hand sidebar to also display the task description and a timer, allowing participants to track the remaining time without distraction, and a button to show the sample application stimulus.

We also integrated a set of convenience features that were common across the tools we analyzed. These included the ability to roll back the application to previous iterations, and a live stream of the model’s output to provide immediate feedback that the agent was actively working. In our implementation, the live stream was deliberately blurred, preventing participants from actually reading or copying code while still conveying progress (see the bottom left in Figure~\ref{fig:ourtool}).

Once the tasks were finalized (see Section~\ref{task-design}), we evaluated system performance across different LLMs from different vendors. In our tests through our custom-built platform, Claude Sonnet 4 consistently outperformed other models on the study tasks, delivering faster responses while maintaining high output quality. On this basis, we selected Anthropic Claude Sonnet 4 (20250514 release) as the model underpinning our platform. For implementation, we opted for a single self-contained HTML file including JavaScript and CSS, simplifying deployment and minimizing overhead. To further improve throughput, we introduced a diff-based update mechanism: instead of rewriting the entire file, the model was instructed to return only the code fragments that required modification in response to user prompts.

\begin{figure}[t!]
    \centering
    \fbox{\includegraphics[width=\linewidth]{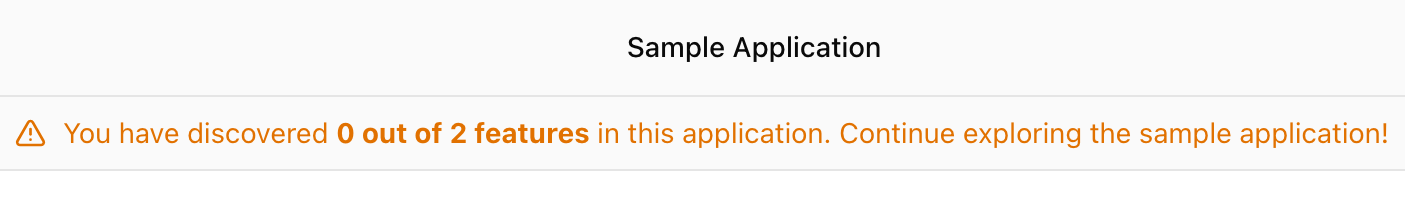}}
    \caption{While exploring the sample application, participants are provided with feedback about the number of relevant features they have interacted with.}
    \Description[Screenshot of feature tracking feedback]{The screenshot shows a window titled "Sample Application." A warning icon and message in orange text state "You have discovered 0 out of 2 features in this application. Continue exploring the sample application!" in a banner at the top of the window.}
    \label{fig:feature-tracking}
\end{figure}

We ensured that the sample applications presented to participants during the study can be replicated fully by participants by also using our own tool to build the sample applications according to the task specifications (see Section \ref{section:tasks}). While reviewing the sample applications, participants further benefited from a feature-tracking mechanism: We instrumented the sample applications with additional code manually after construction through the vibe coding platform to track if and when a feature is used. We then displayed feedback to the user to indicate if they tried all features already, as can be seen in Figure~\ref{fig:feature-tracking}. This information was also logged for further analysis.

\section{Pilot Study}
\label{section:pilot}

We conducted a pilot evaluation with 10 students who were recruited from the same pool as the main body of participants. The primary objective of the pilot evaluation was to determine a sound time limit for each of the three assessment instruments and to test our vibe coding platform. The pilot study had the same structure as our main experiment; all participants filled out the demographic survey first, after which they filled out the questionnaires in random order and then solved the vibe coding tasks on our platform. To limit learning effects, half of the participants, selected at random, submitted the vibe coding tasks first before they were asked to respond to the questionnaires. However, unlike the main study, the pilot essay prompt had a word limit of 150 words.

For the SCS1 12-item subset, the results indicated a high internal consistency: Cronbach's alpha $= 0.83$, Feldt $95\%$ CI $= [0.62,0.95]$, meaning that even the lower confidence bound (0.62) is above the minimum acceptable threshold (0.6) for exploratory scales~\cite{hair2019mda}. 9 of 10 participants completed all 12 items, suggesting only minimal evidence of speededness, and 96\% of all questions were answered. For ICAR16, we gave the participants a 16-minute timer in the pilot test. 8 of 10 participants reached their last item (placing the test right at the 80\% threshold for speededness~\cite{cintron2021methods}) and 90\% of all tasks (144/160) were answered. Only 3 participants needed the full timer; 7 terminated the survey before the timer expired. For the essay task, 9 participants wrote full essays; one participant terminated the task after only a few minutes and after writing only 39 words.

For the SCS1 12-item subset, we retained the pilot time limit for the main study. Because this is the first administration of this instrument with an explicit timer, and because the pilot showed strong internal consistency and minimal speededness (9/10 finished; 96\% of items answered), we believed that altering the limit would introduce unnecessary variance. For ICAR16, we reduced the limit from 16 to 12 minutes, as this remains within common ICAR practice (10-16 minutes) and most participants finished before the 16-minute cap (7/10). Finally, for the essay task, we raised the target length to 300-450 words (as 150 words was much too short for substantive responses) and extended the time to 20 minutes. Additionally, for all surveys, we decided that participants would not have the option of exiting each assessment early.

\begin{figure*}[t!]
    \centering
    \includegraphics[width=0.8\linewidth]{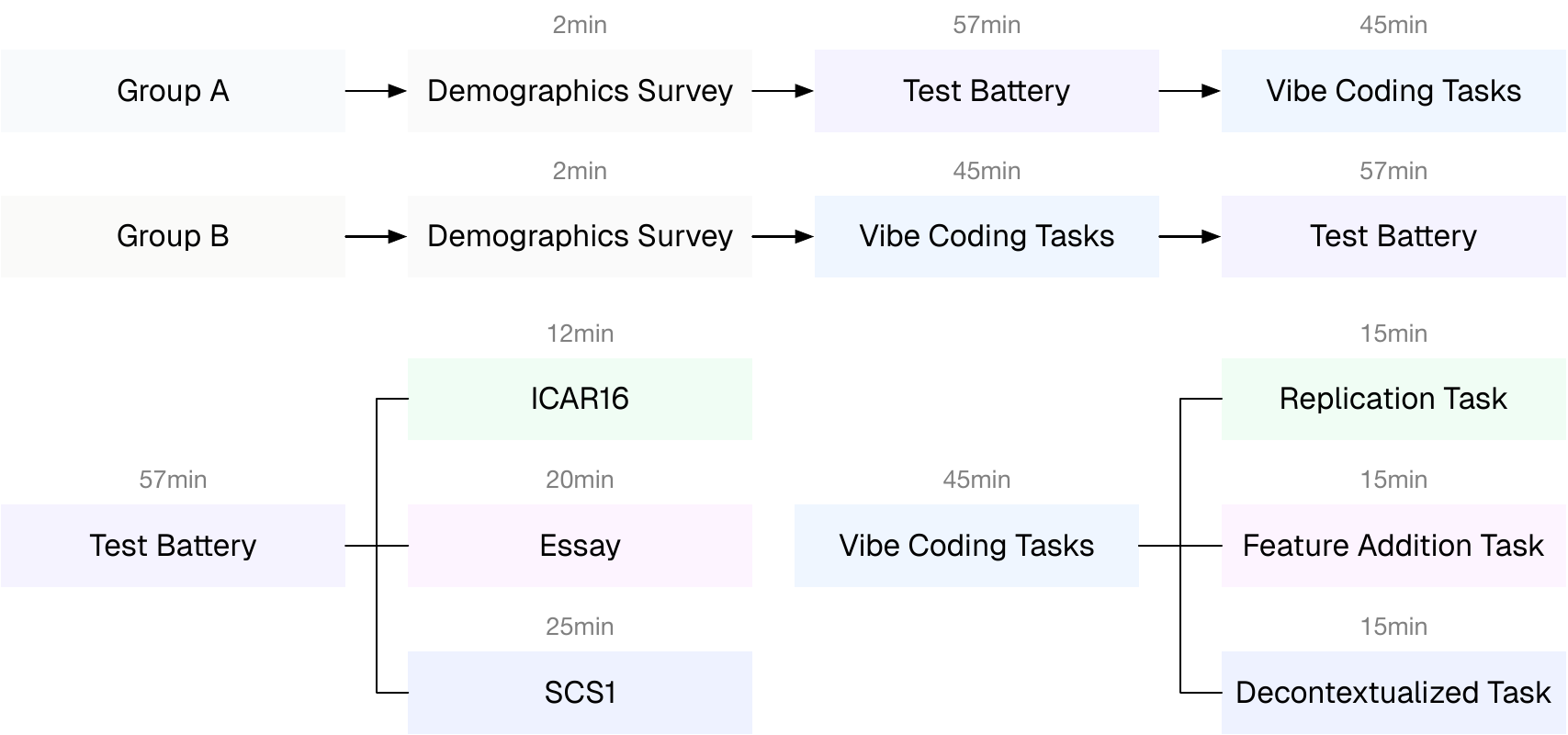}
    \caption{A diagram depicting the flow of our experiment. Participants were randomly divided into Group A and Group B. The order of the instruments in the test battery and the vibe coding tasks was randomized for each participant.}
    \label{fig:experimentflow}
    \Description[Flowchart of the study]{Flowchart of the study procedure with two counterbalanced orders. Group A completes the Demographics Survey (2 min), then the Test Battery (57 min), then the Vibe Coding Tasks (45 min). Group B completes the Demographics Survey (2 min), then the Vibe Coding Tasks (45 min), then the Test Battery (57 min). The Test Battery consists of three instruments totaling 57 min: ICAR16 (12 min), Essay (20 min), and SCS1 (25 min). The Vibe Coding Tasks consist of three 15 min tasks: Replication Task, Feature Addition Task, and Decontextualized Task. Within the Test Battery and within the Vibe Coding Tasks, the order of their components is randomized for each participant.}
\end{figure*}

While analyzing student solutions on the vibe coding platform, we observed that some participants were unaware of base-application features relevant to the tasks they were asked to implement or extend. To preserve construct validity, i.e., the ability to vibe code and not the ability to understand how an existing application works, we updated the platform so that users were explicitly notified of untested features in the sample application (in a small banner at the top of the sample application window, see Figure~\ref{fig:feature-tracking}).

\section{Method}

\subsection{Objectives}
We preregistered three research questions about the role of written communication skills for vibe coding:

\begin{enumerate}
  \item \textbf{RQ1: Task-type specificity.} 
  Does written-communication proficiency relate more strongly to performance on the decontextualized task than to performance on the two context-rich tasks (replication and feature addition)?
  
  \textit{H1.} The association between written-communication proficiency and performance will be larger for the decontextualized task than for each context-rich task.

  \item \textbf{RQ2: Overall association and independence from general reasoning.} 
  (a) Are written-communication skills positively correlated with vibe coding performance? 
  (b) Does this correlation remain after statistically controlling for domain-general cognitive skills?
  
  \textit{H2a.} Written-communication skills will correlate positively with average performance across the three tasks.\\
  \textit{H2b.} No directional hypothesis is specified regarding the partial association after controlling for domain-general reasoning.

  \item \textbf{RQ3: Relative predictive value.} 
  In predicting vibe-coding task performance, how do written-communication skills compare with CS achievement?
  
  \textit{H3.} No directional hypothesis is specified about which predictor will be larger.
\end{enumerate}

\subsection{Participants}

After receiving ethics approval from our institution, ETH Zurich, we conducted a cross-sectional laboratory study with the help of the Decision Science Laboratory (DeSciL) at ETH. DeSciL maintains a large volunteer pool of students from ETH and the University of Zurich, from which we recruited our participants. Each participant received a flat cash payment of 55 CHF for a single session that lasted 1 hour and 45 minutes in the DeSciL controlled computer laboratory. To participate in the study, the students had to have completed an introductory course in computer science, have prior experience working with LLMs for solving computer programming tasks, and meet at least a C1-level of proficiency in English according to the Common European Framework of Reference for Languages~\cite{council2001common}. We chose to recruit 100 participants to achieve sufficient power to at least calculate an overall association between the constructs; a priori power analysis for Pearson’s $r$ indicated that a sample of $N$=85 would be sufficient to detect a medium-sized association ($r$=0.30) with 80\% power in a two-tailed test at $\alpha$=0.05. 

After signing consent forms, the participants completed a brief demographic survey on their age, gender, and field of study. We also asked participants how often they used large language models and the extent to which they perceived them as helpful in daily life. Participants then completed vibe coding tasks and a battery of survey instruments in counterbalanced order (see Figure~\ref{fig:experimentflow}).

\subsection{Analysis Plan}

The primary measures in our study are (i) written-communication skills, (ii) CS achievement (SCS1 12-item subset score), (iii) domain-general cognitive skills (ICAR16 scores) and (iv) vibe coding performance. All of these measures are normalized to a $[0,1]$ range. We answered our research questions in the following way:

\begin{enumerate}
  \item \textbf{RQ1}: We implement a mixed-effects model to test whether written-communication skills relate differently to performance across vibe task types, with the Writing x Task-Type interaction as the primary confirmatory test ($\alpha$ = .05). In contrast to our preregistration, we first compare the correlation between the task types and written communication before deciding whether to construct the model.
  \item \textbf{RQ2a}: We compute the Pearson correlation between written-communication proficiency and vibe coding performance, reporting $r$, a 95\% confidence interval (Fisher $z$), and a two-tailed $p$ value.
  \item \textbf{RQ2b}: We fit an ordinary-least-squares (OLS) regression of aggregate performance on written-communication proficiency and ICAR16 (both z-scored). We report the writing coefficient with 95\% CI, standardized beta, and semi-partial $R^2$ for writing.
  \item \textbf{RQ3}: We conduct hierarchical OLS regressions to compare unique variance: (a) We enter the CS achievement score, then add writing and record $\Delta R^2_{\text{write}|\text{CS}}$, (b) Reverse the order to obtain $\Delta R^2_{\text{CS}|\text{write}}$. We compare the two $\Delta R^2$ values to describe which predictor adds more unique variance; standardized final-step coefficients with 95\% CIs are also reported.
\end{enumerate}

For the grading of the written communication assessment, we refer to the previous sections. For the grading of the vibe coding tasks, one member of the author team manually and blindly reviewed each application and scored each core feature (as defined on the predefined grading rubric) on a 4-point Likert scale, with anchors defined as \emph{not acceptable}, \emph{below expectations}, \emph{meets expectations}, and \emph{exceeds expectations}, as proposed in the literature \cite{mkpojiogu2017can}. 

\section{Results}

\subsection{Overview}

Of the 100 students who participated in the study, the demographic survey indicated that there were 56 women, 44 men, and no agender or nonbinary participants. The average age of 99 of those participants was 25.0 (SD = 3.5); one participant did not enter a number when asked about their age. 51 participants were engineering and technology students, 20 were from the natural sciences, 11 were from the social sciences and the humanities, and the remaining students were from other fields. Our system logs show that for all but 3 of the 300 vibe coding task submissions, participants had explored all features in the sample application before submission.

After grading the 100 essays, our essay graders reached an ICC(2,2) consensus value of 0.735. In line with our preregistration plan, this prompted a round of recalibration and rescoring, as the value was below the 0.750 threshold. We asked the two graders independently to take another look at their own grades and verify if any should be regraded to maintain consistency with their other grades; this lowered the ICC(2,2) value to 0.731. At this point, we terminated the process, as future rounds seemed unlikely to make a difference, and because the value we arrived at is above the widely cited acceptability cutoff of 0.7 for reliability coefficients~\cite{nunnally1994psychometric} (and above the ``good'' cutoff of 0.6 for ICC in particular~\cite{hallgren2012computing}). After this, 21\% of the essays differed by >10\% of the grading scale, and for those essays, the difference was resolved by our third grader as we specified in our plan. The descriptive statistics for the essay grades and the other primary constructs are shown in Table~\ref{tab:instrument_scores}.
    
\begin{table}[t!]
\centering
\begin{tabular}{lcc}
\toprule
\textbf{Instrument} & \textbf{Mean (SD)} & \textbf{Min--Max} \\
\midrule
SCS1 & 0.53 (0.24) & 0.08--1.00 \\
ICAR16 & 0.56 (0.18) & 0.19--1.00 \\
Writing Skills & 0.72 (0.11) & 0.30--0.97 \\
\addlinespace
Vibe Coding Tasks & 0.45 (0.19) & 0.00--0.83 \\
\quad Replication Task & 0.46 (0.32) & 0.00--1.00 \\
\quad Feature Addition Task & 0.55 (0.23) & 0.00--1.00 \\
\quad Decontextualized Task & 0.34 (0.26) & 0.00--0.83 \\
\bottomrule
\end{tabular}
\caption{Descriptive statistics for the results from our study instruments, showing mean, standard deviation (SD), and the minimum and maximum scores achieved. All scores were normalized to [0, 1], with 1 denoting the maximum score possible.}
\label{tab:instrument_scores}
\end{table}

Cronbach's alpha for the SCS1 grades was 0.734, which is above Nunnally et al.'s acceptability threshold, but Cronbach's alpha for ICAR16 was 0.625, which is below the 0.7 threshold but above the threshold of 0.6 for exploratory research~\cite{hair2019mda}. Because of this, we used the ICAR16 measure primarily as a covariate and did not analyze ICAR16 item-type subscales such as verbal reasoning.

\newcommand{\CorrCell}[1]{\num{#1}}

\begin{table}[t!]
\centering
\begin{subtable}{\linewidth}
  \centering
  \begin{tabular}{lrrr}
    \toprule
             & CS & Cog. & Writing \\
    \midrule
    Cog.       & \CorrCell{0.4172} \pval{<.001} & \cellcolor{gray!8} & \cellcolor{gray!8} \\
    Writing  & \CorrCell{0.1256} \pval{.213} & \CorrCell{0.3648} \pval{<.001} & \cellcolor{gray!8} \\
    Vibe     & \CorrCell{0.3861} \pval{<.001} & \CorrCell{0.3522} \pval{<.001} & \CorrCell{0.2902} \pval{.003} \\
    \bottomrule
  \end{tabular}
  \caption{Pearson correlations}
\end{subtable}
\vspace{4mm}
\begin{subtable}{\linewidth}
  \centering
  \begin{tabular}{lrr}
    \toprule
             & CS & Writing \\
    \midrule
    Writing  & \CorrCell{-0.0315} \pval{.757} & \cellcolor{gray!8} \\
    Vibe     & \CorrCell{0.2812} \pval{.005} & \CorrCell{0.1856} \pval{.066} \\
    \bottomrule
  \end{tabular}
  \caption{Partial correlations controlling for cognitive ability}
\end{subtable}
\caption{Correlation matrices between CS achievement (CS), cognitive ability (Cog.), writing skills (Writing), and vibe coding performance (Vibe). All values are Pearson's $r$ with two-tailed \textit{p}-values in parentheses. Partial correlations (b) control for cognitive ability.}
\label{tab:correlations}
\end{table}

We computed the zero-order correlations between the four primary constructs we measured (Table~\ref{tab:correlations}-a). All correlations were positive and statistically significant ($p < 0.05$), except for the correlation between CS achievement and writing skills, which was small and non-significant. In particular, both writing skills and CS achievement correlate with performance on vibe coding. After controlling for performance on the domain-general cognitive test (Table~\ref{tab:correlations}-b), the correlation between CS achievement and vibe coding skills remained significant (Pearson's $r = 0.281$, $p = 0.005 < 0.05$), but not the correlation between written communication and vibe coding skills (Pearson's $r = 0.186$, $p = 0.066 > 0.05$). 

\subsection{Research Questions}

For \textbf{RQ1}, we asked if written-communication skills relate more strongly to performance on the decontextualized task than to performance on the two context-rich tasks. A simple check showed that we could not reject the null hypothesis, as the correlation (Pearson's $r$) between writing skills and the decontextualized task (0.239) was lower than for the feature addition task (0.245; see Table~\ref{tab:essay_corr}).

\begin{table}[t!]
\centering
\begin{tabular}{l r r}
\toprule
Task & \multicolumn{1}{l}{$r$} & \multicolumn{1}{l}{$p$} \\
\midrule
Context-Rich Task 1 (Replication)   & 0.154 & 0.126 \\
Context-Rich Task 2 (Feature Addition) & 0.245 & 0.014 \\
Decontextualized Task & 0.239 & 0.017 \\
\bottomrule
\end{tabular}
\caption{Pearson correlations between the three vibe coding task scores and writing performance (two-tailed $p$). As specified in Table~\ref{tab:correlations}, the overall Pearson's $r$ between the task average and writing was 0.290.}
\label{tab:essay_corr}
\end{table}

For \textbf{RQ2}, we asked whether written-communication skills are positively correlated with vibe coding performance (a relationship that we hypothesized exists) and whether this correlation persists after controlling for domain-general cognitive skills, for which we made no hypothesis. In accordance with our preregistration plan, we computed Pearson's $r$ for the former and found a correlation of $r = 0.290, p = 0.003$, 95\% confidence interval of $[0.099, 0.460]$, indicating a significant but small ($<0.3$) positive correlation, so we could reject our null hypothesis. This correlation becomes non-significant when controlling for the ICAR16 results: partial $r = 0.186$,  95\% confidence interval of $[-0.012, 0.386]$, $p = 0.066$.

For \textbf{RQ3}, we asked how the predictive value of CS achievement and written-communication skills compare on vibe-coding performance. We compared the unique variance in aggregate vibe-coding accuracy explained by written-communication proficiency and CS achievement using hierarchical OLS, per the preregistered plan. Entering the CS achievement score first and then adding writing increased $R^2$ from $0.150$ to $0.208$ ($\Delta R^2_{\text{write}\mid\text{CS}}=0.059$). Reversing the order, entering writing first and then adding SCS1 increased $R^2$ from $0.083$ to $0.208$ ($\Delta R^2_{\text{CS}\mid\text{write}}=0.125$). Thus, CS achievement scores contributed roughly twice the unique variance of writing scores in predicting aggregate performance, although writing still added nontrivial incremental variance. Using both predictors in the final model, standardized coefficients indicated positive associations for both: Writing had the parameter $\beta=0.244$ \,[95\% CI $0.063$, $0.425$], $p=0.009$; and CS had the parameter $\beta=0.356$ \,[95\% CI $0.176$, $0.537$], $p<0.001$. Descriptively, CS achievement is the stronger predictor of aggregate vibe-coding accuracy, but written-communication skills also show a reliable unique contribution beyond CS achievement.

\section{Exploratory Analysis}

As the outcomes on the CS achievement test and vibe coding tasks were significantly and moderately positively correlated, and the correlation remained significant even when controlling for domain-general cognitive skills, we analyzed the secondary data to explore what could contribute to this relationship. One exploratory hypothesis that we developed was that the effect was moderated by experience and interest; namely, that students with more CS experience and hence greater CS achievement might be more familiar with how to work with large language models (LLMs). To investigate this, we computed how the participants' self-reported Likert-scale frequency of LLM usage (encoded as a 1-5 variable, with 5 mapped to the highest frequency) correlated with the vibe coding performance. To our surprise, we found a significant negative correlation ($r = -0.258; p = 0.010$). We also found a significant negative correlation between frequency of LLM usage and written communication skills ($r=-0.282, p=0.005$), but no correlation between LLM usage and CS achievement ($r=0.001, p=0.994$) or LLM usage and the ICAR16 results ($r=-0.070, p=0.486$).

A second exploratory question that we considered was whether the writing assessment primed participants to write better prompts for the vibe coding tasks. To analyze that, we computed a two-sided independent-samples Student's t-test on the vibe coding performance of students who began the study with the survey battery versus those who began the study with the vibe coding tasks. We found no evidence of a difference; the means were $0.446$ and $0.448$, respectively, with $p=0.953$. 

Our third exploratory question, designed to better understand the robustness of our results, was on how the quality of the user prompts correlated with the essay grades and performance on our vibe coding tasks. To measure the prompt quality, we followed a two-step approach:

\begin{enumerate}

\item First, we organized a human evaluation of the writing quality of the prompts. We asked the creators of our essay grading rubric to design a grading rubric for the writing quality of prompts for human-LLM interaction. They designed a rubric (included in our supplementary materials) that involved coherence, task-appropriate complexity, and instructional clarity, bearing in mind that prompts tend to be expressed in the imperative grammatical mood, and that many elements of essay writing (such as sentence flow) are less relevant in this context. We then recruited a member of the vibe coding expert panel to grade the prompts using this rubric, who graded the full prompt sequence for each task attempt as a single unit. The rubric was designed by a team that had not seen our prompt data. The expert panelist, who was not involved in any previous rounds of grading, graded the prompts blind to all other participant data, including essay scores, task outcomes, and the generated code.

\item Second, to triangulate the results from the human evaluation, we considered the \emph{lexical richness} of the prompts, a term ``also known as lexical diversity''~\cite{yang2023predicting}, which has been found to be significantly correlated with writing quality~\cite{yang2023predicting,xie2015study,woods2023multi}. Many automatic measurements of lexical richness exist; some measures involve counting the frequency of uncommon words in the text (which is called lexical sophistication), while others quantify the balance between vocabulary variety and text length. However, lexical sophistication may not be suitable or meaningful in our context; first, it would require an appropriate reference corpus, and second, prior work has found that when substituting individual words in human-generated prompts with synonyms, the rarity of the synonym does not correlate with improved output quality from the LLM~\cite{leidinger2023language}. Instead, following McCarthy and Jarvis~\cite{mccarthy2010mtld}, we chose two common measures that are length-independent and applicable to short texts; the HD-D, a probability-based measure that captures how much vocabulary variety appears across random word samples drawn from the text, and the MTLD (Measure of Textual Lexical Diversity), which measures how long, on average, the text sustains varied vocabulary before words start repeating. Both measures have been validated as robust alternatives to simpler metrics like type-token ratio, which is known to be confounded by text length~\cite{mccarthy2010mtld}. We used the \texttt{lexical-diversity} library in Python to calculate this.

\end{enumerate}

\begin{table}[t]
\centering
\begin{tabular}{lrrr}
  \toprule
   & \multicolumn{1}{l}{Essay grade} & \multicolumn{1}{l}{Essay LD} & \multicolumn{1}{l}{Vibe perf.} \\
  \midrule
  HD-D & \CorrCell{0.2059} \pval{.0399} & \CorrCell{0.2080} \pval{.0379} & \CorrCell{0.3096} \pval{.0017} \\
  MTLD & \CorrCell{0.2480} \pval{.0129} & \CorrCell{0.2580} \pval{.0095} & \CorrCell{0.3429} \pval{.0005} \\
  \bottomrule
\end{tabular}
\caption{Pearson's $r$ between prompt-related measures and outcome variables. HD-D and MTLD measure the lexical diversity (LD) of students' prompts. Essay LD refers to the lexical diversity of the student's essay.}
\label{tab:prompt-correlations}
\end{table}

We then computed how a student's average prompt quality measure correlated with other metrics. All correlation coefficients were positive and significant; students who exhibited varied vocabulary in their prompts were also significantly more likely to (i) exhibit varied vocabulary in their essays, (ii) score a higher grade on their essay, and (iii) achieve a greater performance on the vibe coding tasks (see Table~\ref{tab:prompt-correlations}). Students who exhibited higher human-graded prompt quality were also significantly more likely to have a higher grade on their essays and to score better on the vibe coding tasks (Pearson's $r = .353$ and $r = .479$, respectively, with $p < .001$). 

To examine whether the observed associations are consistent with prompt quality serving as a linking mechanism between writing skill and vibe coding performance, we used the human-graded prompt quality to conduct mediation analysis, which is a statistical technique that has sometimes been applied in CHI research papers~\cite{yurrita2025towards,cho2019}. Following Hayes' textbook on the subject~\cite{hayes2022introduction}, we used 10{,}000 bootstrap resamples. As we noted before, writing skill was associated with prompt quality (path $a = 0.35$, $p < .001$), and prompt quality was associated with vibe coding performance when controlling for writing skill (path $b = 0.43, p < .001$). The total association between writing and vibe coding (path $c = 0.29$, $p = .003$) was reduced to non-significance when prompt quality was included in the model (path $c' = 0.14$, $p = .145$). The indirect effect was 0.152, with a 95\% bootstrap confidence interval of $[0.061, 0.279]$ that excluded zero, and accounted for approximately 52\% of the total association. These results are consistent with prompt quality serving as an intermediate variable linking writing proficiency to task performance: participants with stronger writing skills produced clearer, better-organized prompts, which were in turn associated with more successful outcomes. As we conducted a cross-sectional study, we do not wish to draw causal conclusions; rather, this pattern provides response-process evidence supporting the construct validity of our vibe coding measure. Rather than capturing some ability unrelated to prompting, the behavioral outcomes appear to reflect differences in how effectively participants specified their intent to the LLM.

\section{Discussion}

We believe that the primary and most interesting finding from our study is that both written communication skills and CS achievement contribute positively, significantly, and independently to GUI-oriented vibe coding performance, and in the case of CS achievement, in a way that cannot be explained by domain-general cognitive ability alone. Furthermore, our exploratory analysis suggests that this relationship is not explained either by high-performing CS students having had greater experience in working with LLMs, and because the study was designed in such a way that the code generated from the LLM was not visible (only the rendered output itself), we know that CS students had no obvious advantage in interpreting the model output. As our study is correlational, we cannot make causal claims about the impact of CS achievement on vibe coding; such claims could only be investigated in controlled studies. On the other hand, our study results can be used to generate hypotheses for causal mechanisms that could be explored in such studies. For instance, it is possible that the structured thinking that students learn in computer science classrooms lends itself well to specifying program behavior; vibe coding may draw, in some sense, on core CS knowledge or parts of the ``hidden curriculum'' of CS, such as problem decomposition and algorithmic thinking, which we believe is a part of CS achievement. It is also possible that a greater command of CS-related vocabulary can help students write better prompts.

As for written communication skills, we believe that students with high scores on the construct are not only skilled at clear writing, but also at following conventions for structured writing. Such students may therefore also be skilled at structuring their prompts in a way that helps them accomplish specific tasks. In the current educational environment, where U.S. 4th grade reading scores are still well below pre-pandemic levels~\cite{nces2024naepreading, nces2024naepreadingweb}, and both PISA scores in reading performance and adult literacy scores are trending downwards~\cite{OECD2023,OECD2024SurveyAdultSkillsUS}, it is plausible that any positive impact of language training on students' ability to convert their ideas into code could motivate students to learn to express their ideas more clearly.

The implications of our study could also inform tool design. A significant number of tools are devoted to teaching LLM users to write better prompts; for instance, ROPE~\cite{Ma2025}, which trains users to write effective prompts by providing automatic feedback on prompt clarity; Wu et al.'s tool~\cite{wu2022ai}, which automatically elaborates based on the LLM’s response to the original prompt; and Arawjo et al.'s system~\cite{arawjo2024chainforge} for testing and evaluating prompts. In our view, these systems implicitly assume that the main bottleneck is users' ability to express precise requirements in natural language, which is an assumption that our study lends tentative support for.

Apart from this central finding, we sought to understand how written communication skills impact certain broad types of vibe coding tasks, in particular the difference between non-contextualized tasks and context-rich tasks. Before the study took place, we believed that written communication skills were more important for the former, as the LLM agent could not be relied on to fill in any semantic gaps in students' specifications for completely unfamiliar applications. However, we found no evidence for this hypothesis in our study. Future studies could investigate how other types of vibe coding tasks besides context-rich and decontextualized tasks are impacted by student skills and abilities. Such studies may also investigate broader populations than students and how they perform on different vibe coding tasks.

While our work analyzed how prompt quality and the lexical diversity of participants' prompts relates to and mediates the association between writing proficiency and GUI-oriented vibe-coding performance, future work could categorize prompting styles (e.g., verbose versus curt, iterative workflows versus one-big-prompts) and examine how student skills correlate with such workflows and how it affects their vibe coding performance, although we recommend that any such studies consider explicitly beforehand how the results could inform pedagogical practice or tool design. 

Lastly, we find it interesting that prior LLM usage is negatively and significantly correlated with written communication and vibe coding performance. We speculate that this relationship is caused either by (i) LLMs having a negative effect on students' ability to express themselves or (ii) that students who are less proficient at writing are more likely to use LLMs, or (iii) a combination of both. However, the exploratory analysis did not show evidence for LLM usage correlating either negatively nor positively with CS achievement. Other studies have also been ambivalent on the effects of AI exposure on CS and computer programming outcomes; for instance, Kazemitabaar et al.'s controlled study from 2023~\cite{kazemitabaar2023studying} found no statistically significant difference in performance on programming retention tests between learners who had access to AI code generators during training and those who did not.

\subsection{Result Generalizability}

Our study is constrained by its focus on GUI-oriented vibe coding. To probe how far our findings might extend to more algorithmic forms of vibe coding, we conducted an exploratory analysis of the most computation-focused component of our task suite. In the meal-planning task, participants were given a catalog of meals annotated with attributes such as cost, calories, protein, and sugar, and asked to implement behavior that assembled a weekend meal plan under budget and nutritional constraints. This required specifying logic that iterated over the meal dataset, aggregated numeric fields (for example, total and average calories and protein), and compared those aggregates against target thresholds to compute summary statistics and whether they met the given threshold. It is plausible that the performance on this component is an indicator of a participant's ability to compute a simple algorithm via natural-language prompts, as opposed to focusing solely on surface-level GUI actions.

We found that CS achievement and writing skills were also significant predictors of performance on this data-centric aspect of the meal-planning task. CS achievement showed a Pearson correlation of $r = 0.320$ ($p = 0.001$), and writing skills showed a Pearson correlation of $r = 0.202$ ($p = 0.044$). These effect sizes are comparable to, though slightly smaller than, the associations we observed for overall vibe-coding performance. Taken together, they provide tentative evidence that the pattern we observe is not limited to only surface-level GUI construction: both CS knowledge and written communication proficiency also appear to matter when vibe coding requires participants to elicit nontrivial algorithmic behavior from the model. At the same time, this analysis is based on a single task in a specific domain, and should therefore be interpreted cautiously. A fuller assessment of generalizability will require future work that systematically varies task type and includes a broader range of data-centric and algorithm-focused vibe-coding scenarios.

Our results are also constrained by our focus on ``no-code'' or ``pure'' vibe coding. In our setting, participants could not inspect or edit the underlying source, so CS knowledge could only support performance indirectly (e.g., through problem decomposition or mental models of control flow and state), rather than through direct code manipulation. As such, our estimates for the role of CS achievement in general LLM-augmented programming should be viewed as a lower bound: in more permissive workflows, the same underlying skills can contribute both via prompt quality and via targeted edits to generated code. At the same time, introducing code access is unlikely to simply scale up the role of CS and scale down the role of writing in a linear way. When both channels are available, users must choose how to distribute work between natural-language specification and code editing; this opens up new strategies (e.g., using prompts for high-level restructuring and code edits for local fixes) and may produce interaction effects between skills. Future work should examine how these skills shape strategy choice and outcomes in mixed-mode environments where reading and editing code are also on the table.

\subsection{Limitations and Threats to Validity}

Our study has several limitations and threats to validity. First, we note that the vibe coding construct can cover a broader category of tasks than those we studied, for example data analysis. It is possible that our results would be different if we had examined such tasks. Second, our measure of written communication skills, while developed and evaluated by five experts (four of whom worked independently), has not been previously deployed. This might impact its construct validity. 

The ecological validity of the study is impacted by the controlled laboratory environment we used, which differs from most real-world vibe-coding settings. In the study, we analyzed ``no-code'' or ``pure'' vibe coding only, meaning that participants could not review the code generated, and the results may not translate directly to more permissive LLM-augmented programming workflows. Similarly, all tests and tasks were strictly timed, and we observed in the vibe coding component that several participants ran out of time when they were close to arriving at a correct solution. Task framing may have also affected participants' natural strategies. It is possible that the results were also impacted by students misunderstanding the tasks they were meant to implement, although task misreading and comprehension skills is likely an unavoidable source of construct-irrelevant variance in this context.

Lastly, we note that our sample consisted of university students. This limits the generalizability of our findings to other populations, such as professional developers or citizen programmers, who may have different vibe-coding workflows than our participants.

\section{Conclusion}

We presented a study on the impact of measured computer-science achievement, domain-general cognitive ability and written communication skills on the ability to create GUI-oriented applications using a vibe coding approach in which the code is hidden from view. We found that both computer science achievement and written communication skills were positively and significantly correlated with vibe coding skills ($r = .39$ and $r = .29$, respectively), a correlation that persists in the former case after controlling for cognitive ability. Furthermore, both constructs contribute independently to the GUI-oriented vibe coding performance. Our results encourage investigation into the hypothesis that computer science instruction can lead to better outcomes on LLM-guided programming, which is a question that should be investigated in future controlled studies.

\begin{acks}

We thank Dr. Kimberly Lewis and David Camorani from the Language Center of UZH and ETH Zurich for the essay grading rubric. We also thank Kimberly and Dr. Giorgio Iemmolo for the help with the prompt analysis in our exploratory analysis. Any flaws in our work, or in the way that we incorporated their advice, remain our own. We thank the anonymous members of our expert panel and the Decision Science Laboratory at ETH Zurich, and extend special thanks to Dr. Haukur Thorgeirsson, Sandra Wiklander, Anita Barr, and Dr. Tracy Ewen.

\end{acks}

\bibliographystyle{ACM-Reference-Format}
\bibliography{BIBLIO}

\end{document}